\documentclass[lettersize,journal]{IEEEtran}
\usepackage{amsmath,amsfonts}
\usepackage{algorithm}
\usepackage{array}
\usepackage[caption=false,font=normalsize,labelfont=sf,textfont=sf]{subfig}
\usepackage{textcomp}
\usepackage{stfloats}
\usepackage{url}
\usepackage{enumerate}
\usepackage{verbatim}
\usepackage{graphicx}
\usepackage{xcolor}
\usepackage{cite}
\usepackage{multirow}
\usepackage{amsmath}
\usepackage{algorithm}
\usepackage{algpseudocode} 
\usepackage{tabularx,booktabs,textcomp}
\usepackage{amsthm}

\usepackage{amsmath,amssymb,amsthm}

\begin{document}

\title{A Comprehensive Metric for Resilience Evaluation of Power Distribution Systems under Cyber Attacks}

\author{{Sagar Babu Mitikiri, K. Victor Sam Moses Babu, Vedantham Lakshmi Srinivas, Pratyush Chakraborty, Mayukha Pal$^{*}$}

\thanks{$^{*}$(Corresponding author: Mayukha Pal)}

\thanks{Mr. Sagar Babu Mitikiri is a Data Science Research Intern at ABB Ability Innovation Center, Hyderabad 500084, India, and also a Research Scholar at the Department of Electrical Engineering, Indian Institute of Technology (ISM), Dhanbad 826004, IN.}
\thanks{Mr. K. Victor Sam Moses Babu is a Data Science Research Intern at ABB Ability Innovation Center, Hyderabad 500084, India and also a Research Scholar at the Department of Electrical and Electronics Engineering, BITS Pilani Hyderabad Campus, Hyderabad 500078, IN.}
\thanks{Dr. Vedantham Lakshmi Srinivas is an Asst. Professor with the Department of Electrical Engineering, Indian Institute of Technology (ISM), Dhanbad 826004, IN.}
\thanks{Dr. Pratyush Chakraborty is an Asst. Professor with the Department of Electrical and Electronics Engineering, BITS Pilani Hyderabad Campus, Hyderabad 500078, IN.}
\thanks{Dr. Mayukha Pal is with ABB Ability Innovation Center, Hyderabad-500084, IN, working as Global R\&D Leader – Cloud \& Analytics (e-mail: mayukha.pal@in.abb.com).}
}

\maketitle

\begin{abstract}
     Power distribution systems (PDS) serve as the backbone of our modern society, ensuring electricity reaches homes, businesses, and critical infrastructure. However, the increasing digitization and interconnectivity of these systems have exposed them to cyber threats. This study presents a comprehensive approach to evaluate and enhance the resilience of PDS under cyber attacks using the Common Vulnerability Scoring System (CVSS) and complex network parameters. By systematically assessing vulnerabilities and computing resilience once critical CVSS thresholds are reached, this work identifies key resilience metrics including the critical loads service requirements. The proposed methodology improves system resilience through strategic tie-line switching, which is validated on the modified IEEE 33-bus system. Four case studies are conducted, illustrating the performance of the proposed methodology under various cyber attack scenarios. The results demonstrate the effectiveness of the approach in quantifying and enhancing resilience, offering a valuable tool for PDS operators to mitigate risks and ensure continuous service delivery to critical loads during the exploitation of cyber threats.
\end{abstract} 

\begin{IEEEkeywords}
Resilience, Power Distribution Systems, Critical Loads, CVSS, CVE IDs, Cyber attacks.
\end{IEEEkeywords}

\section{Introduction}
\label{section: Introduction}


\subsection{Background and Motivation}

In modern power distribution systems, it is necessary to keep critical loads operational even in the presence of contingencies. The idea of resilience is a relatively different concept when compared to the reliability that has been studied in \cite{rostami2023reliability}. According to the IEEE Task Force, resilience for critical infrastructure is defined as the capacity to withstand and minimize the duration and impact of disruptive events such as its ability to anticipate, absorb, adapt to, and recover rapidly from such events \cite{stankovic2018definition}. The focus on maintaining critical operations implies that even during any events or disruptions, the essential functions should be restored immediately and remain functional. For power systems, it refers to ensuring that critical infrastructures like hospitals, data centers, etc., must remain operational. But it has several definitions given by several US and UK organizations like the National Renewable Energy Laboratory (NREL), Department of Energy (DoE), Sandia National Laboratories, and Electric Power Research Institute (EPRI). The detailed definitions of resilience from these organizations are present in \cite{dwivedi2024technological}. 

The resilience of the distribution systems is affected by various natural hazards like hurricanes, cyclones, earthquakes, and also due to unnatural activities like cyber attacks \cite{umunnakwe2021quantitative, panteli2015grid, goldenberg2001recent}. Many researchers have done a lot of work in quantifying, evaluating, and improving resilience due to natural events \cite{dwivedi2024technological}. Advances in digitalization and industrial automation have led to an increase in the number of digital and communication devices in the power distribution system making it a Cyber Physical Power System (CPPS). Considering this enhancement in the integration of Information and Communication Technology (ICT) in CPPS brings the concern of cyber threats vulnerabilities that arise due to various vulnerabilities present in the physical or ICT infrastructure \cite{venkataramanan2019measuring}. It is necessary to model both the physical and cyber components of the CPS to fully understand the concepts of resilience. Cyber attacks are designed to disrupt the operation of critical loads by attempting to disconnect them from generation sources. Cyber attacks play a crucial role in impacting the resilience of Distribution systems and critical loads to degrade the system performance or cause long-term damage by disrupting their normal operations.

With the increased penetration of advanced infrastructure equipment, telecommunication, and control system technologies, the threat of cyber intrusions and sabotage in power distribution systems has become a significant concern. These cyber attacks can result in severe damage and may also lead to blackouts. As nearly all industrial operations and everyday human activities are heavily dependent on electricity, such disruptions could lead to substantial economic losses and impact critical infrastructure, including military and other sensitive operations. Consequently, enhancing resilience against both physical and cyber-attacks is crucial. It is essential to prevent unauthorized access and destructive actions within intelligent power systems to ensure the continued provision of reliable and sustainable energy \cite{ghanbari2023resilient}.  

The disruptions may occur due to the destructive error made by the operator depending on the compromised sensor measurements or by remote operation of the grid by malicious intruder. Factors like financial benefits, large blackouts, or a combination of both motivate the intruders to initiate cyber attacks on the system. There are many reported incidents of cyber attacks in power distribution systems. Such as the Ukrainian Power grid attacks in 2015 that disrupted power disruptions across multiple regions affecting 225000 customers \cite{case2016analysis}. In 2016, the same grid was affected but it occurred on a comparatively small scale \cite{conway2016analysis}. IN 2019, US power utility companies were affected by spear-pishing campaigns that aimed to control systems access gaining. Even though there were no disruptions, it evolved the threat of cyber attacks on critical infrastructure \cite{kumar2022apt}. The cyber attacks that affect resilience could take many forms. Some of the common attacks are False Data Injection Attacks (FDIA), Denial of Service (DoS), Man in the Middle (MITM), insertion of malware, and physical damage, etc., FDIA includes the injection of fake data in a communication line between the control center ad field sensors to deceive the operators leading to energy theft, miscalculation of local marginal prices (LMP), for illegal profits, and physical damage depending upon the intruders goal. DoS attacks are realized as attacks on the routing signal that lead to congestion and delays in communication channels, resulting in jamming and compromising electronic devices. These attacks generally restrict the access of services and resources through communication flooding with unwanted signal traffic. Insertion of malware/worms are of different types,  


\subsection{Literature Review}

Before implementing resilience measures, it is crucial to quantify the resilience of a system numerically, allowing operators to quickly assess the system's changing operating conditions \cite{wang2015research}. Several works have presented on the improvement and enhancement of resilience due to natural hazards through various approaches like complex network theory, data driven, and technological advances. The data driven approaches consist of predictive risk analytics such as weather based risk assessment, power outage estimations, line fault detections, and various sensor data driven approaches etc. Complex network theory is utilized in obtaining the connectivity loss, reliability, and distribution, cascading effects, blackout size, etc. Technical advancements constitute the resilience enhancement through mobile energy storage systems \cite{wang2018resilience}, peer-to-peer energy trading \cite{babu2024resilient}, additional switches, and renewable energy integration.

The power distribution system is the most vulnerable part of the electric power delivery process. Its radial configuration complicates the restoration efforts since it typically lacks alternative pathways for rerouting power when disruptions occur. Factors further affecting the resilience of PDS include the increasing occurrence of extreme weather events, which can severely damage the infrastructure, the aging of the existing infrastructure, and the growing integration of cyber components used for automation, monitoring, and control, which are particularly susceptible to remote cyber-attacks \cite{kandaperumal2020resilience}. The effect of cyber attacks on the power system components has already been studied \cite{dagoumas2019assessing, chen2020study}. However, the impact of the attacks is not well defined due to the importance of security issues in the critical infrastructure of applications like SCADA (Supervisory Control and Data Acquisition Systems) \cite{cruz2016cybersecurity}. The resilience enhancement for cyber attacks by exploitation of these securities in the network requires a dedicated metric CVSS \cite{cvss_v3_1}. An analysis tool for enabling microgrid resilience is proposed in \cite{venkataramanan2019cyphyr} and implemented in two phases the planning and operation phases making the operator take proactive measurements for resilient operation. A novel microgrid scheduling model is introduced that ensures resilience against risks associated with failures, load uncertainties, and variations in renewable generation. \cite{liu2020robust}. The resilience impact factors are integrated with a fuzzy Choquet integral in \cite{venkataramanan2019cp} for security assessment of cyber physical microgrids. In \cite{venkataramanan2020cp}, a cyber physical resilience metric is proposed considering the multi criteria decision making. 

\subsection{Contributions and Organization}

The review of the existing literature indicates that research on evaluating the resilience of CPPS under cyber attacks is relatively limited compared to the extensive studies focused on power system resilience under extreme weather conditions. To address this gap, this article proposes a resilience evaluation method based on complex network parameters, specifically designed to ensure the continued service of all critical loads during cyber attacks.

The main contributions of this article are summarized as follows:

\begin{itemize}
    \item A comprehensive evaluation of the vulnerabilities is performed in the power distribution systems using CVSS, providing a detailed understanding of the system's exposure to cyber threats using the Common Vulnerability and Exposure (CVE) database.
    \item The resilience in power distribution systems is assessed by ensuring that all critical loads are met. Thus this proposed approach prioritizes maintaining essential services during cyber attacks.
    \item The incorporation of complex network parameters in quantitative evaluation of resilience, understanding the system's robustness in terms of critical load serviceability.
    \item The resilience of the power distribution system is improved by strategically closing appropriate tie line switches. This method effectively strengthens the system's ability to withstand and recover from cyber attacks.
    \item The proposed methodology is verified with the modified IEEE 33-Bus system, which is particularly vulnerable to disruptions due to its radial nature. This evaluation is conducted across four case studies, each considering cyber attacks at different locations within the system.
\end{itemize}

The methods used to measure the resilience, complex network parameters, CVSS metrics, and electrical parameters are presented in Section \ref{Section: Methodology}. The results of the proposed method validating with the various case studies are discussed in Section \ref{Section: Results and Discussions}. Finally, Section \ref{Section: Conclusion} draws the conclusions of the work done. 

\section{Methodology}
\label{Section: Methodology}

\subsection{Complex network parameters for EDS}
The complex network theory has been utilized in evaluating the resilience of various critical infrastructures. The complex network theory applications for resilience evaluation were initially proposed by Réka Albert and Albert-László Barabási. In the power systems field, various problems related to reliability and resilience have been solved using complex network theory including, critical node identification, restoration strategies, connectivity loss, damage and improvements, unserved energy/loads, blackout size, and cascading effects. 

This evaluation of resilience using complex network parameters is composed of modeling the electrical distribution system as a graph. This graph, represented as $G= (N, E)$ is a collection of $N$ number of nodes/vertices and $E$ edges, where each edge is connected from node i to j ($i,j=1,2,..., N$). Various topological parameters are computed to the obtained network for evaluating the resilience. The following are the network parameters used for evaluating the resilience:

\subsubsection{Algebraic Connectivity} This metric is a measure of the structural resilience of the network and its ability to withstand the faults caused by the disconnectivity of the graph due to natural events or cyber-attacks. Algebraic connectivity is defined as the second-largest eigenvalue of the difference of the Laplacian matrix derived from the degree matrix of the corresponding network and its adjacency matrix. This parameter is denoted as $\lambda_{2}$, for a disconnected graph its value will be $0$ and $0 < \lambda_{2} \leq N$ for a connected graph. The higher the value of $\lambda_{2}$, the more resilient the network will be when it is partitioned into isolated regions \cite{molloy1995critical}. 

\subsubsection{Average Shortest Path} The average shortest path is a measure of the average distance between all pairs of vertices in the graph. It is also known as characteristic path length and denoted by $L$. It is defined as the average of the shortest path lengths between all pairs of vertices in the graph \cite{albert2002statistical}. It is calculated as follows:
\begin{equation}
    L = \frac{2}{N(N-1)} \sum_{u \neq v} d(u,v)
    \label{Eqn:Avg shortest path}
\end{equation}
Where $d(u,v)$ represents the shortest distance path between the vertices $u$ and $v$. Lower values of $L$ indicate that most of the nodes/vertices are easily connected to any other nodes in the graph within fewer steps. It provides the efficiency of how quickly power or information is transmitted between the nodes.

\subsubsection{Average Betweenness Centrality} In complex network analysis, centrality measures are used to identify the key nodes in the network, which are critical for structural and behavior analysis. These are of various types, of which Betweenness Centrality $(C_B)$ is one. It uses the number of shortest paths that travel through a vertex to determine how influential it is \cite{chen2014assessing}. A vertex that regularly serves as a bridge along the shortest paths between other vertices is said to have a high $C_B$ value. For a given vertex $(v)$ is computed as:
\begin{equation}
    C_B(v) = \sum_{\substack{s,t \in V \\ s \neq t \neq v}} \frac{\sigma_{st}(v)}{\sigma_{st}}
    \label{Eqn:Betweenness centrality}
\end{equation}
Where $\sigma_{st}$ indicates the overall number of shortest paths between the vertices $s$ and $t$ and $\sigma_{st}(v)$ is the number of shortest paths passing through the vertex $v$. The above summation is considered for all distinct vertices pairs $s$ and $t$, excluding the vertex $v$ in the graph.

The overall idea behind the Average Betweenness Centrality (C$_{AB}$) on average how central the vertices are \cite{gago2014betweenness, barthelemy2004betweenness}. It is computed by considering the means of $C_B$ values for all vertices in the graph and is as follows:
\begin{equation}
    C_{AB} = \frac{1}{n} \sum_{v \in V} C_B(v)
    \label{Eqn:Avg betweenness centrality}
\end{equation}

\subsubsection{Diameter} The diameter of the graph is defined as the maximum eccentricity of any vertex in the graph. In other words, it represents the longest shortest path between any pair of vertices in the graph \cite{hernandez2011classification}. The diameter of the graph is denoted as $D_g$ and is computed as follows
\begin{equation}
    D = \max_{u,v} \, d(u,v)
    \label{Eqn:Diameter}
\end{equation}

\begin{figure*}
    \centering
    \includegraphics[scale=0.5]{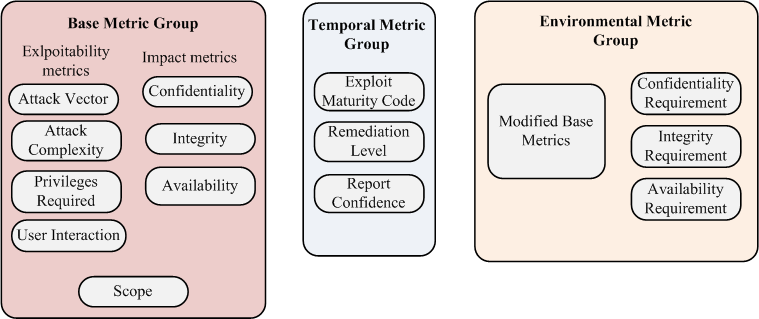}
    \caption{CVSS Metrics}
    \label{fig:CVSS metrics} 
\end{figure*}

\subsection{Cyber security parameters}

Common Vulnerability Scoring System (CVSS) is an open framework for communicating the characteristics and severity of software vulnerabilities \cite{cvss_v3_1}. It captures the principal technical characteristics of software, hardware, and firmware vulnerabilities and outputs the severity of vulnerability relative to others as numerical scores. CVSS is composed of three metric groups: Base, Temporal, and Environmental. The classification of the metric groups of CVSS is shown in Fig. \ref{fig:CVSS metrics}. The benefits of CVSS are standardized vendor provision and a platform for challenging vulnerability scoring methodology.

\subsubsection{Base Metrics}
This metric group indicates the inherent qualities of vulnerabilities that hold constant throughout time and in user environments. This scoring system assumes that the attacker has gained access to the target system's weaknesses. It is composed of two metrics: Exploitability and Impact metrics. 

The exploitability metrics reflect the characteristics of the vulnerable components. These exploitability metrics are further divided into four groups which are scored relative to the vulnerable component properties that lead to a successful attack. The subgroup of metrics are as follows:

\begin{itemize}
    \item \textbf{Attack Vector (AV):} It reflects the context by which vulnerability exploitation is possible. Its value will be larger when the attack has occurred more remotely. It takes values such as Network (N), Adjacent (A), local (L), and Physical (P).
    \item \textbf{Attack complexity (AC):} It explains the external requirements that must be completed for the attacker to make use of the vulnerability. Such circumstances might necessitate computing exceptions or the gathering of additional target information, as explained below. Crucially, the evaluation of this metric does not include any conditions on user engagement to take advantage of the vulnerability. These metrics take higher values for the least complex attacks. It takes values like Low (L) for the repeatable attacks and High (H) for the successfully launched attacks that go beyond the attacker's control. 
    \item \textbf{Priviliges Required (PR):} It describes the range of privileges before exploring the vulnerability, for an attacker to possess. The list of possible values for this metric is None (N) for no access, Low (L) for the attacker who has access to only non-sensitive information, and High (H) for significant access before the attack.
    \item \textbf{User Interaction (UI):} It describes the need for a manual user except the attacker. When no user interaction is required for the task of successfully compromising the vulnerable component, then this base score is greater. It takes values of None (N) for no interaction and Required (R) if any user is required before exploitation.
\end{itemize}
\subsubsection{Scope} This metric captures whether the vulnerability in one component impacts the resources of other components beyond its scope. If a vulnerability in one component affects the other component which is in a different security scope then the scope changes. The base score is also the greatest only when a scope changes. The possible values of the scope metric are: Unchanged (U) when the affected and exploited components are from the same scope region and Changed (C) for different regions.

\subsubsection{Impact Metrics:} 
Impact metrics measure the consequences of a successfully exploited vulnerability on the component that experiences the most severe outcome directly linked to the attack. Analysts should limit these impacts to a plausible and definitive result that they are confident an attacker could realistically achieve. Impact metrics consider only privileges gained, accessibility increase, and any negative outcomes after exploitation while scoring. If a scope change occurs, then the impact metrics should reflect either the vulnerable or impacted component (which suffers more), whereas if the scope doesn't change then the impact metrics should be reflected to the vulnerable component. The impact metrics are further divided as follows:

\begin{itemize}
    \item Confidentiality (C): It is defined as limiting access to the disclosures and information only to authorized personnel. It measures the confidentiality of the information resources due to successful exploitation. When the impacted component suffers a severe loss, then its value will be greatest. It takes values like High (H) if there is complete confidentiality loss, Low (L) when the attacker gains the restricted information but, doesn't have control over it, and Non (N) for no confidentiality loss.
    \item Integrity (I): It refers veracity and trustworthiness of information. Its base score is the highest when the consequence to the impacted component is greater. The possible values for this metric are High (H) for total loss of integrity, Low (L) when modification of data is possible, but control over the consequence of modification is not possible by the attacker, and None (N) when no integrity loss.
    \item Availability (A): This metric refers to the information resource accessibility such as attack network bandwidth, processor cycles, and the impacted component availability. It measures the availability of the impacted component after successful exploitation. The possible values for this metric are High (H) for the total availability loss, Low (L) for reduced performance or possible repeated exploitations, and None (N) for no impact on availability.
\end{itemize}

\subsubsection{Other CVSS Metrics}The other CVSS metrics are Temporal and Environmental. Temporal metrics reflect the features of vulnerability that vary with time but not across the environments. Environmental metrics refer to the vulnerability characteristics that are related to a particular environment of the user. A detailed explanation of these metric groups is provided in the CVSS documentation \cite{cvss_v3_1} provided by the FIRST organization.


\subsection{Metrics Scoring using CVSS} 
When the Base metrics are assigned, a Base score ranging from 0 to 10 is computed. The base equation is a combination of both the Exploitability and Impact subscores that are derived from their respective metric groups. The base score thus obtained is adjusted by incorporating the Temporal and Environmental metric groups for a better representation of relative vulnerable severity within a user's environment at a given time. While scoring these metrics is optional, and is recommended only for obtaining accurate scores. The computation of the overall CVSS Base metric score (CVSS$_{Base}$) is as follows:

\begin{algorithm}[h]
\caption{CVSS Base Score Calculation Algorithm}
\label{Algorithm: CVSS Base Score}
\begin{algorithmic}[1]

\State \textbf{Compute Impact Sub-Score (ISS)}
\[
ISS =  1-\left[(1-\text{C})\times(1-\text{I})\times(1-\text{A})\right]
\]
\If{Scope is Unchanged}
    \State $Impact = 6.42\times ISS$
\Else
    \State $Impact = 7.52 \times(ISS - 0.029) -3.25 \times (ISS - 0.02)^{15}$
\EndIf

\State \textbf{Compute Exploitability:} 
\[
\text{Exploitability} = 8.22 \times \text{AV}\times\text{AC}\times\text{PR}\times\text{UI}
\]
\If{Impact $\leq$ 0}
    \State BaseScore = 0
\Else
    \If{Scope = Unchanged}
        \State $CVSS_{Base} = \text{Roundup}(\min[\text{Impact} + \text{Exploitability}, 10])$
    \Else
        \State $CVSS_{Base} = \text{Roundup}(\min[1.08 \times (\text{Impact} + \text{Exploitability}), 10])$
    \EndIf
\EndIf
\end{algorithmic}
\end{algorithm}

\subsection{Electrical Service Parameter}
\subsubsection{Critical Load Served ($E_{CL}$)} It is a quantitative parameter. It measures the amount of critical load served during the outage due to cyber attacks \cite{gao2016resilience}. It is defined as the ratio of the amount of critical load served during the cyber attack outage event to the total critical load present in the system. Its value ranges from 0 to 1. Mathematically, it is formulated as:
\begin{equation}
    E_{CL} = \frac{\text{Critical load served}}{\text{Critical load demand}}
\end{equation}

\subsection{Overall Resilience Score} The resilience score ($R$) for the proposed is computed as a combination of the normalized scores of complex network parameters, electrical parameters, and the CVSS base score. Since the standard resilience value ranges from 0 to 1. The parameters are complemented and normalized to the same scale to make the above-mentioned parameters suitable for this. Since the CVSS is measured on a scale of 0 to 10, where 10 indicates high severity and 0 for No severity, this metric is

\subsection{Proposed Resilience Evaluation Method}


The resilience score is computed based on the severity of CVSS$_{Base}$ score. The severity score was obtained from the Common Vulnerability and Exposure (CVE) database. The CVE-IDs \cite{christey2007vulnerability} present in the database of CVSS are used to identify the specific vulnerabilities that are being scored through the Algorithm \ref{Algorithm: CVSS Base Score}. Thus obtained score is compared with the severity scale provided by the FIRST organization as shown in Table \ref{tab:severity scale}. 

\begin{table}[]
    \caption{CVSS severity Sclae}
    \label{tab:severity scale}
    \centering
    \begin{tabular}{c c}
         \hline
         \textbf{Score}&  \textbf{Rating} \\
         \hline
         0 & None \\
         \hline
         0.1 - 3.9 & Low \\
         \hline
         4 - 6.9 & Medium \\
         \hline
         7 - 8.9 & High \\
         \hline
         9 - 10 & Critical\\
         \hline
    \end{tabular}
\end{table}

\begin{equation}
    R = f(\lambda_2, L, C_{AB}, D, E_{CL})\\
    \label{Eqn:Basic resilience}
\end{equation}

\begin{equation}
    R = g(\lambda_2, L^{-1}, C_{AB}^{-1}, D^{-1}, E_{CL})\\
    \label{Eqn:Modified Basic resilience}
\end{equation}

IF the CVSS severity reaches a high/critical level, then the resilience of the system is computed by using the above-mentioned parameters. The resilience score is described as a function ($f$) of five metrics out of which four are complex network parameters ($L$, $\lambda_2$, $C_{AB}$, $D$) and one is the electrical parameter (E$_{CL}$) as shown in Eq. \ref{Eqn:Basic resilience}. For a system to be more resilient, the comparative value of the parameters $\lambda_2$, and E$_{CL}$ should be more and the remaining parameters should be less for different cases. Hence to make the resilience score values between 0 and 1, the inverse values of $L$, $C_{AB}$, and $D$ are used and are represented as $L^{-1}$, $C_{AB}^{-1}$, and $D^{-1}$ respectively. After this modification, the resilience score is computed through Eq. \ref{Eqn:Modified Basic resilience}. Where g is a function, that sums up the normalized values of all the parameters. The calculation of the resilience score is as follows:

\begin{equation}
    R = \Hat{\lambda_2} + \Hat{L^{-1}} + \hat{C_{AB}^{-1}} + \hat{D^{-1}} + \hat{E_{CL}}\\
    \label{Eqn:Final resilience eauation}
\end{equation}

Where $\hat{\lambda_2}$, $\hat{L^{-1}}$, $\hat{C_{AB}^{-1}}$, $\hat{D^{-1}}$, and $\hat{E_{CL}}$ are the normalized values for the parameters $\lambda_2$, $L^{-1}$, $C_{AB}^{-1}$, $D^{-1}$, and $E_{CL}$ respectively. These normalized values for each parameter are obtained by dividing them with the respective parameter values when the system is operating in the standard mode as shown in Fig. \ref{fig:Standard}, i.e., when there are no cyber attacks and all the tie line switches are in opened condition.

\begin{figure}[h]
    \centering
    \includegraphics[scale=0.45]{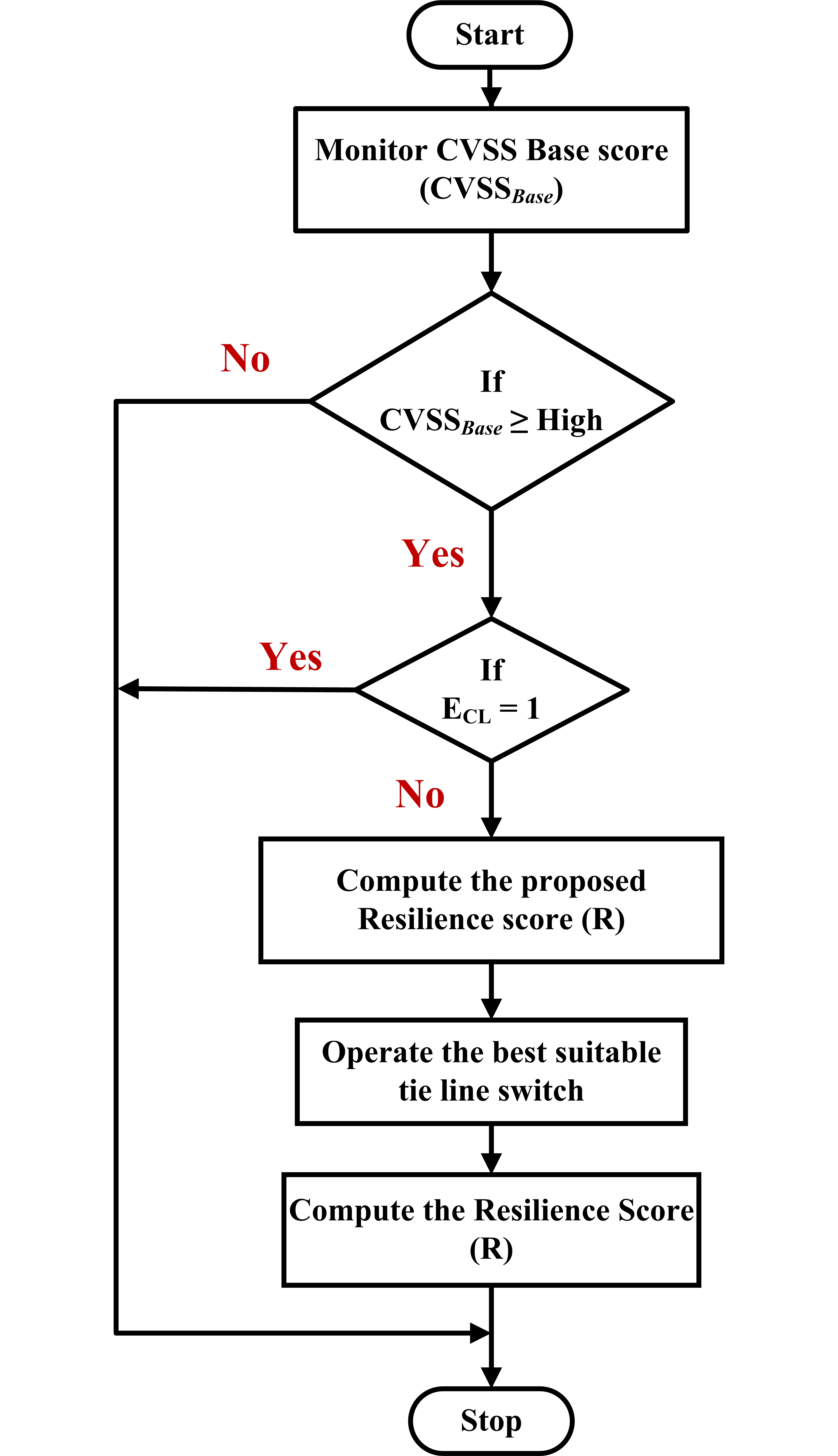}
    \caption{Flowchart for evaluating and enhancing the resilience}
    \label{fig:flowchart}
\end{figure}

The overall methodology of the proposed algorithm is shown in Fig. \ref{fig:flowchart}. Once the CVSS score reaches the severity level the resilience is computed and the appropriate tie line switch is closed to enhance the resilience. After closing the switch, the resilience score for the new system is computed and denoted by $R'$. This methodology deals with the main aim of serving critical loads even during outage conditions.

\section{Results and Discussions}
\label{Section: Results and Discussions}

\begin{figure}[h]
    \centering
    \includegraphics[scale=0.3]{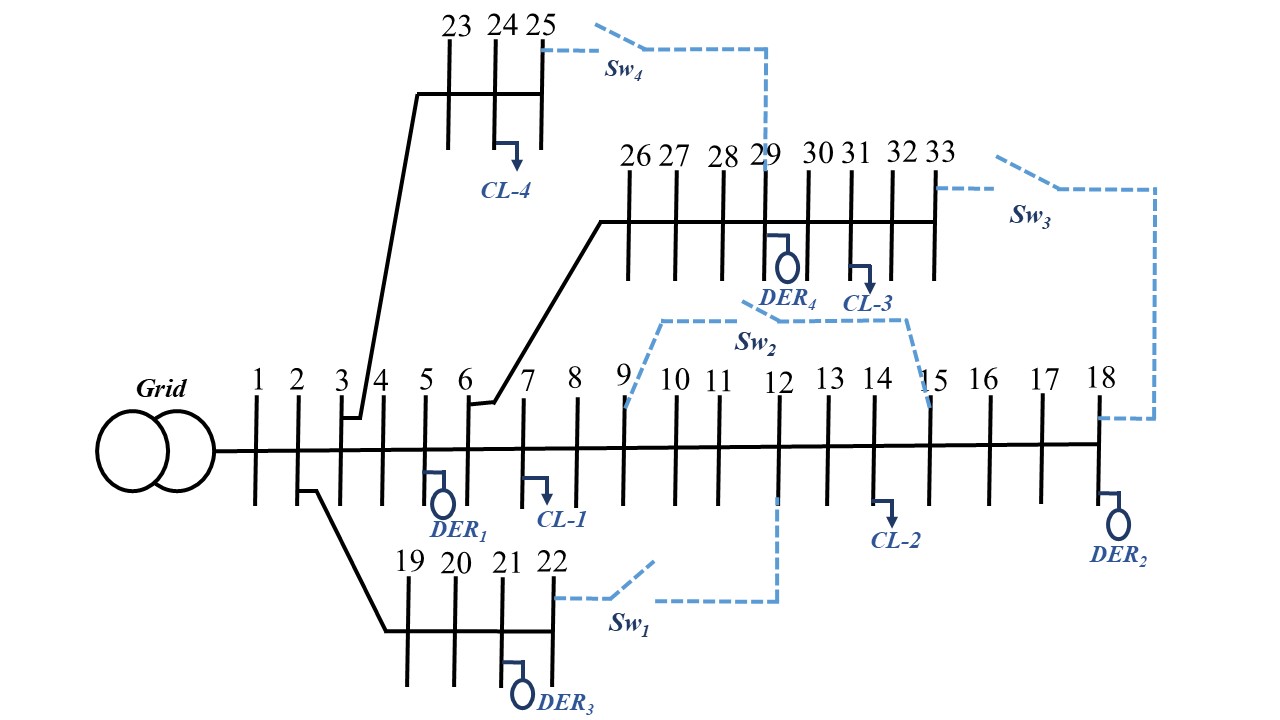}
    \caption{Modified IEEE 33 Bus Distribution System}
    \label{fig:Standard}
\end{figure}

The considered modified IEEE 33 bus test node system is shown in Fig. \ref{fig:Standard}. The Standard IEEE 33 bus distribution system is modified by adding four Distributed Energy Resources (DERs) and four Critical Loads (CLs) to it. Table \ref{tab:DER and Critical Loads} provides the details of these DERs and CLs. The placement of DERs and CLs is done randomly. Four Tie line switches (Sw) are placed for resilience enhancement and to feed the CLs during the event of natural hazards or cyber attacks. Table \ref{tab:switch details} depicts the details of all tie line switches. In normal operating conditions, all the DERs and CLs are connected and the tie line switches are Normally Open (NO) position. 

\begin{table}[h]
    \centering
    \caption{Details of DERs and Critial Loads}
    \label{tab:DER and Critical Loads}
    \begin{tabular}{>{\centering\arraybackslash}p{1cm}>{\centering\arraybackslash}p{2cm}>{\centering\arraybackslash}p{2.5cm}}
         \hline
         \multicolumn{3}{c}{\textbf{DER Details}} \\
         \hline
         \textbf{DERs} & \textbf{Node Location} & \textbf{Ratings}\\
         \hline
         DER-1 & 5 & 720 kW \\
         \hline
         DER-2 & 18 & 800 kW \\
         \hline
         DER-3 & 21 & 760 kW\\
         \hline
         DER-4 & 29 & 800 kW\\
         \hline
         \multicolumn{3}{c}{\textbf{Critical Load Details}} \\
         \hline
         \textbf{CLs} & \textbf{Node Location} & \textbf{Ratings}\\
         \hline
         CL-1 & 7 & 200 kW, 100 kvar\\
         \hline
         CL-2 & 14 & 120 kW, 80 kvar\\
         \hline
         CL-3 & 24 & 1420 kW, 200 kvar\\
         \hline
         CL-4 & 31 & 150 kW, 70 kvar\\
         \hline
    \end{tabular}
\end{table}

\begin{table}[]
    \centering
    \caption{Tie line Switches Details}
    \label{tab:switch details}
    \begin{tabular}{>{\centering\arraybackslash}p{1cm}>{\centering\arraybackslash}p{1cm}}
         \hline
         \textbf{Switch} & \textbf{Location}\\
         \hline
         SW$_1$ & 12 - 21 \\
         \hline
         SW$_2$ & 9 - 15 \\
         \hline
         SW$_3$ & 18 - 33 \\
         \hline
         SW$_4$ & 25 - 29 \\
         \hline
    \end{tabular}
\end{table}


considered system

Figures for the considered system  - keep only one

resilience scores under normal conditions
\subsection{Case-1}

\begin{figure}[h]
    \centering
    \includegraphics[scale=0.3]{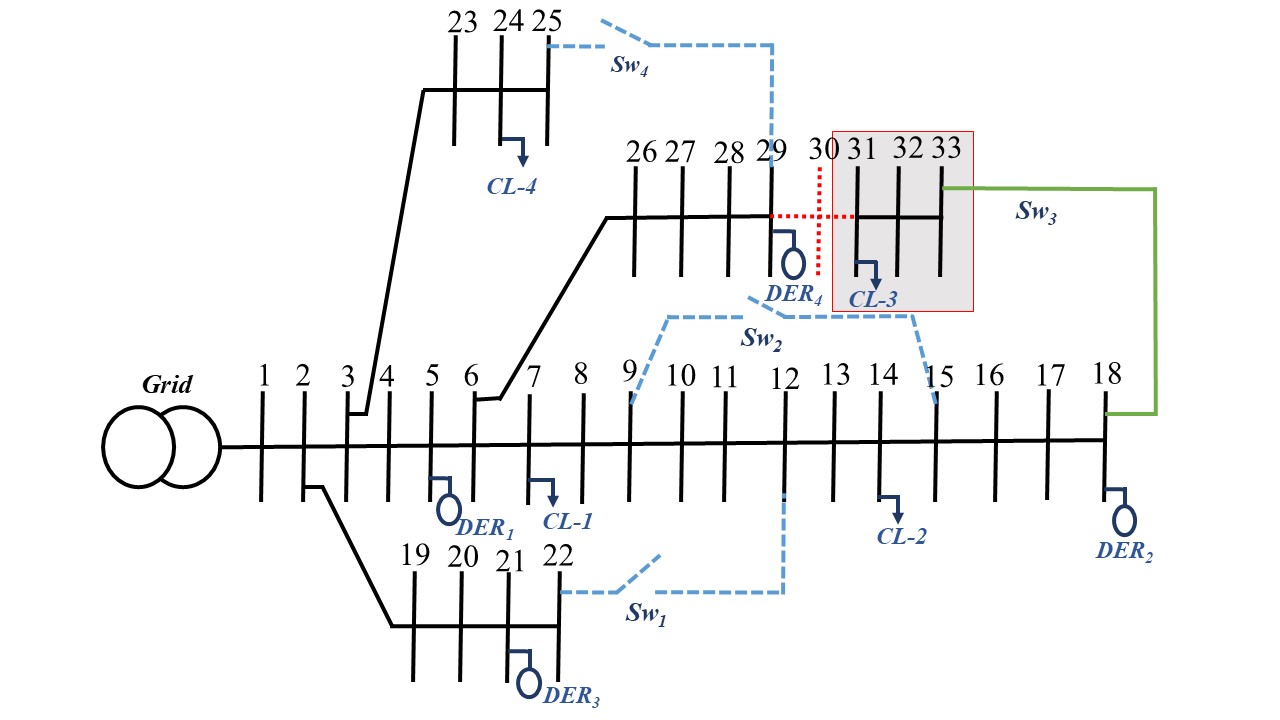}
    \caption{Impact of Node-30 Attack on IEEE 33-Bus System and Resilience Enhancement with SW$_3$ Tie Line Connection}
    \label{fig:case1}
\end{figure}

In this case, a cyber attack is considered to have occurred to disable Node-30, resulting in power disruptions to the subsequent nodes where the CL-3. It is assumed that this attack occurred due to the exploitation of vulnerability with CVE ID: CVE-2020-10937 \cite{CVE-2020-10937} that affects the device operations like disabling the nodes by modifying the system configurations. Once the attack occurs, the entire system after Node-30 won't be able to receive the power, and hence the CL-3 also. As the CVSS$_{Base}$ for this case is High with the numerical value 7.5, the resilience score is computed and found to be 0.31. The resilience of supplying the power to all critical loads was improved by using the timeline switches and in this case, SW$3$ is closed which connects the nodes 18 and 33. The resilience of the system after the closing of the tie line switch was computed and found to be 0.90. Fig. \ref{fig:case1} represents the effect of the attack that occurred on Node-30 after the closure of the switch SW$3$ for serving all the disrupted critical loads.

\subsection{Case-2}

\begin{figure}[h]
    \centering
    \includegraphics[scale=0.3]{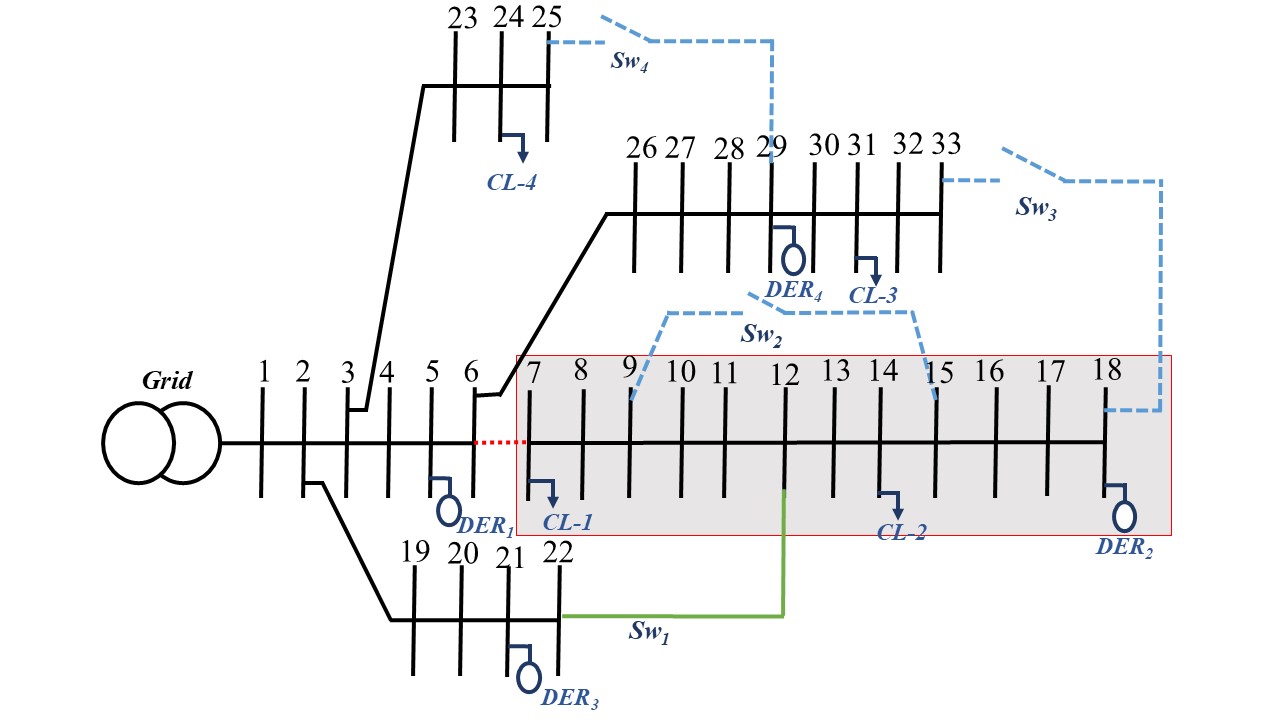}
    \caption{Impact of Attack Disabling Edge Between Nodes 6 and 7 on IEEE 33-Bus System and Resilience Improvement by Connecting Tie Line SW$_1$}
    \label{fig:case2}
\end{figure}

This case considers a cyber attack that has the effect of disabling the power line between nodes 6 and 7 in the system. It is assumed that this attack occurred due to the exploitation of the vulnerability with CVE ID: CVE-2021-40825 \cite{CVE-2021-40825} which exploits system controllers leading to unintended control actions like disconnecting the lines. once the attack is initiated, the nodes after Node-6 won't be able to get power from the grid and the DER-2 present at Node-18 is not capable of supplying power to this region. As the CVSS$_{Base}$ score for this case is 8.6 which is high and the demand for the critical loads CL-1 and CL-2 present in this region are not met with the required amount of energy and hence the resilience drops to 0.75. The resilience was improved by closing the tie line switch SW$_1$ present between nodes 22 and 12. This switch will meet all critical load demands by supplying from both the grid and DER-3. The resilience of the system computed after closing the switch SW$_1$ is 0.90. Fig. \ref{fig:case2} represents the effect of the attack that disconnects the power lines between nodes 6 and 7 after the closure of the SW$_1$ for serving all the disrupted critical loads.

\subsection{Case-3}

\begin{figure}[h]
    \centering
    \includegraphics[scale=0.3]{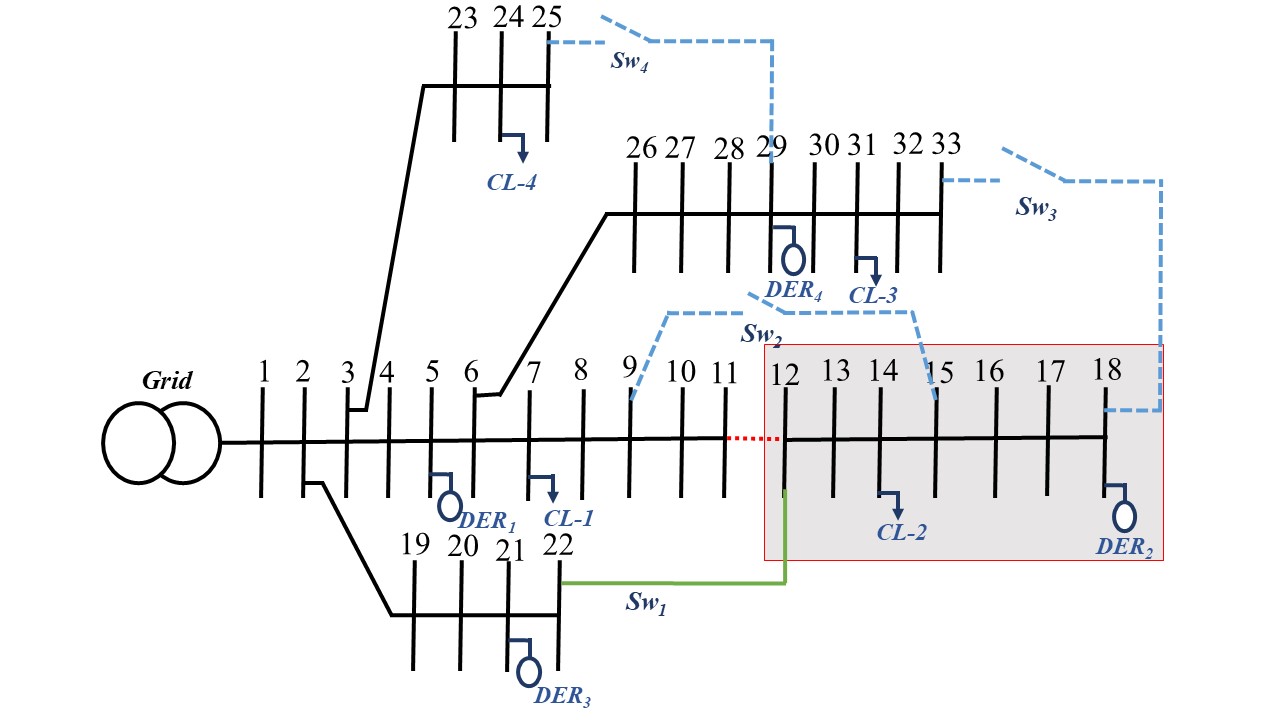}
    \caption{Impact of Attack Disabling Edge Between Nodes 11 and 12 on IEEE 33-Bus System and Resilience Improvement by Connecting Tie Line SW$_1$}
    \label{fig:case3}
\end{figure}

A similar case to the prior is considered here, which is the disabling of power lines between the two nodes 11 and 12. Due to this, the loads from Node 12 to Node 18 including the CL-2 won't be able to get enough supply as the DER-2 present at Node 18 is not capable of meeting the demand. The resilience score for this scenario is obtained as 0.61. The resilience, in this case, is improved by closing the tie line switch SW$1$ and ensuring that enough power is supplied to CL-2. The resilience score computed after closing the switch is obtained as 0.93. Fig. \ref{fig:case3} represents the effect of the attack that disconnects the power line between nodes 11 and 12 after operating the switch SW$_1$ to improve resilience.

\subsection{Case-4}

\begin{figure}[h]
    \centering
    \includegraphics[scale=0.3]{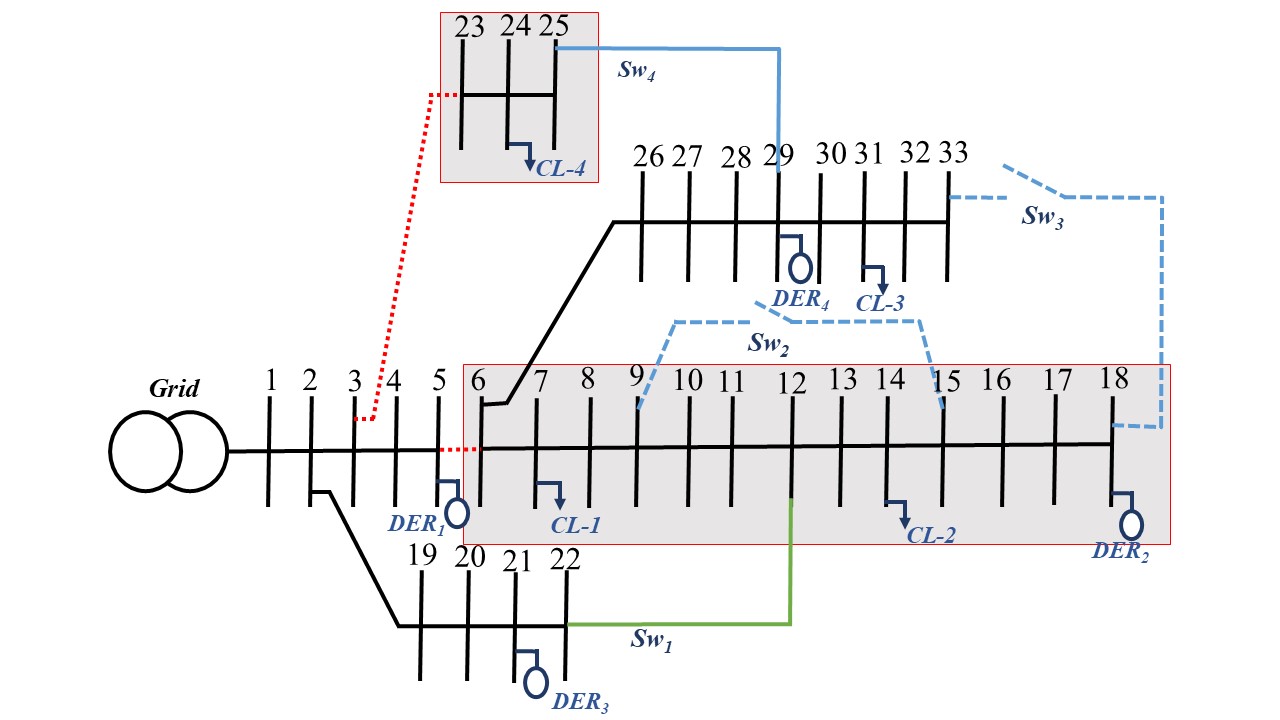}
    \caption{Impact of Cyber Attack on IEEE 33-Bus System by Disabling Edges (3,23) and (5,6), and Resilience Enhancement through Tie Lines SW$_1$ and SW$_4$}
    \label{fig:case4}
\end{figure}

This case considers disconnecting the multiple lines presenting between the node pairs (3, 23) and (5, 6). This scenario is realized by exploiting the vulnerability of CVE ID: CVE-2017-7921 \cite{CVE-2017-7921}, where the attacker may manipulate the commands to circuit breakers, leading to disconnecting multiple lines by affecting the SCADA systems. This attack results in the complete disruption of CL-4 and partial disruption of CL-1 and CL-2 as DER-2 is not capable of supplying power to these loads. In this case, the CVSS$_{Base}$ score is 10 which is critical, and the resilience computed is obtained as 0.55. In this case, the resilience is improved by closing the tie line switches SW$4$ for serving the CL-4 and SW$_1$ for the critical loads CL-1 and CL-2 respectively. The improved resilience score computed is obtained as 0.87. Fig. \ref{fig:case4} represents the effect of the attack that disconnects the power lines between the node pairs (3, 23) and (5, 6) after operating the switches SW$_1$ and SW$_4$ to improve resilience.

\begin{table}[]
    \centering
    \caption{Resilience scores for all the Cases}
    \label{tab:Resilience Scores}
    \begin{tabular}{>{\centering\arraybackslash}p{1cm}>{\centering\arraybackslash}p{2cm}>{\centering\arraybackslash}p{2.2cm}}
         \toprule
         \multirow{2}{*}{\textbf{Case}} & \multicolumn{2}{c}{\textbf{Resilience Scores}}\\
         \cline{2-3}
         & \textbf{During attack} & \textbf{Post Restoration}\\
         \midrule
         Case-1 & 0.31 & 0.90 \\
         Case-2 & 0.75 & 0.91 \\
         Case-3 & 0.61 & 0.93 \\
         Case-4 & 0.55 & 0.87 \\
         \bottomrule
    \end{tabular}
\end{table}

\begin{figure}
    \centering
    \includegraphics[scale=0.35]{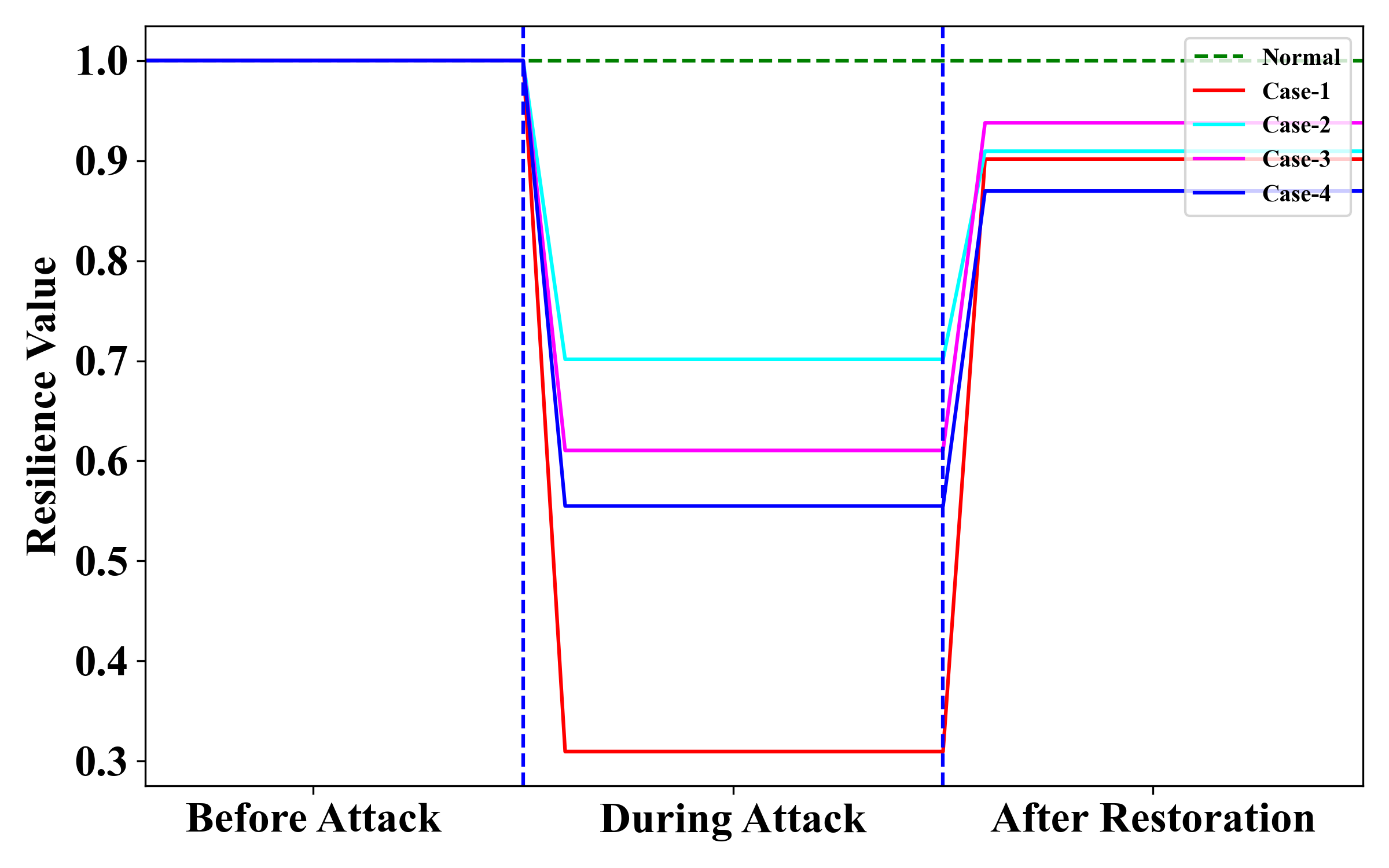}
    \caption{Resilience evaluation curve for all the cases}
    \label{fig:plot}
\end{figure}

Table \ref{tab:Resilience Scores} depicts the resilience scores of all the cases during the attack and after closing the tie line switches. The details of a few more CVE IDs including their descriptions \cite{nvd_search} are shown in Table \ref{tab:CVEID}. Fig. \ref{fig:plot} shows the resilience curves of all the cases for the duration of pre attack, post attack (pre restoration), and the post restoration period. It shows that the initial resilience is equal to the normal conditions. During an attack, the resilience drops drastically leaving most of the critical loads unserved. After the restoration, the value of resilience improves and settles to a value close to the pre attack conditions.

\begin{table}[]
    \centering
    \caption{Descriptional details of CVE IDs }
    \label{tab:CVEID}
    \begin{tabular}{>{\centering\arraybackslash}p{2.3cm}>{\centering\arraybackslash}p{5.5cm}}
       \toprule
       \textbf{CVE ID} & \textbf{Description}\\
       \midrule
       CVE-2015-5371 \& CVE-2018-1002105 & Vulnerability in SCADA systems and firmware allowing remote attackers executing arbitrary code via crafted packets\\
       CVE-2016-6606 & Vulnerability allowing the attackers to execute arbitrary code remotely in Siemens SIMATIC WinCC\\
       CVE-2017-11762 & Vulnerability in Schneider Electric's EcoStruxure control expert leading to unauthorized access and potential system expertise\\
       CVE-2018-12808 & Vulnerability in Modbus protocol implementation in various industrial control systems leading to denial of service\\
       \bottomrule
    \end{tabular}
\end{table}

\section{Conclusion}
\label{Section: Conclusion}
This work introduces a methodology for evaluating and enhancing the resilience of power distribution systems against cyber threats. By employing the CVSS, we quantify vulnerabilities and determine resilience based on the system’s ability to serve critical loads. The resilience metric is further refined using complex network parameters, providing a robust assessment framework. The approach to enhance resilience involves strategically closing tie-line switches, which has shown significant improvement in system resilience as validated on the modified IEEE 33-bus system. The case studies highlight the radial system’s vulnerabilities to cyber attacks at various nodes, demonstrating that strategic network modifications can effectively mitigate these risks. This research contributes a systematic method for the power distribution system's resilience assessment and enhancement, underscoring the importance of integrating cyber-physical considerations into the operational strategies of modern power systems. Future work includes exploring the various cyber attacks that may occur by exploiting the system and validating these cyber attacks across diverse network configurations.

\bibliographystyle{IEEEtran}
\bibliography{main}

\end{document}